# Effects of Bulk and Surface Conductivity on the Performance of CdZnTe Pixel Detectors

Aleksey E. Bolotnikov, C. M. Hubert Chen, Walter R. Cook, Fiona A. Harrison,
Irfan Kuvvetli, and Stephen M. Schindler

California Institute of Technology, Pasadena, CA 91125 USA

*Abstract*—We studied the effects of bulk and surface conductivity on the performance of high-resistivity CdZnTe (CZT) pixel detectors with Pt contacts. We emphasize the difference in mechanisms of the bulk and surface conductivity as indicated by their different temperature behaviors. In addition, the existence of a thin (10-100 A) oxide layer on the surface of CZT, formed during the fabrication process, affects both bulk and surface leakage currents. We demonstrate that the measured *I-V* dependencies of bulk current can be explained by considering the CZT detector as a metal-semiconductor-metal system with two back-to-back Schottky-barrier contacts. The high surface leakage current is apparently due to the presence of a low-resistivity surface layer that has characteristics which differ considerably from those of the bulk material.

This surface layer has a profound effect on the charge collection efficiency in detectors with multi-contact geometry; some fraction of the electric field lines originated on the cathode intersects the surface areas between the pixel contacts where the charge produced by an ionizing particle gets trapped. To overcome this effect we place a grid of thin electrodes between the pixel contacts; when the grid is negatively biased, the strong electric field in the gaps between the pixels forces the electrons landing on the surface to move toward the contacts, preventing the charge loss. We have investigated these effects by using CZT pixel detectors indium bump bonded to a custom-built VLSI readout chip.

## I. INTRODUCTION

Hard X-ray focusing telescopes have been recently proposed for balloon-borne payloads (HEFT [1] and InFocus [2]), and the Constellation X-ray space mission. A position sensitive CdZnTe (CZT) detecting system is currently considered to be the best choice as the focal plane detector for these telescopes. We are developing a system which consists of an array of CZT pixel detectors indium bump-bonded to custom VLSI readout chips. In previous papers [3,4] we have discussed the details of the design and performance of earlier CZT/VLSI hybrids, which have proved that excellent energy resolution (670 eV FWHM at 59.5 keV and −10 C) is possible for such detectors.

We found that bulk and surface conductivity, which determines the total pixel leakage current, can also affect the charge transport in CZT detectors. For example, in pixel detectors with conventional contact geometry; some fraction of the electric field lines originated on the cathode above pixel boundaries terminates at the anode within the gaps between the pixel contacts where the signal charge gets trapped [5-7]. To overcome this effect we place thin steering electrodes (a grid) between the pixel contacts [8,9]. When the grid is negatively biased the strong electric field inside the gaps between the pixels forces the electrons on the surface to move toward the contacts, significantly reducing the charge loss.

The main drawback of using the steering electrodes, as in the case of coplanar grid detectors, is an extra leakage current component that significantly increases the total leakage current per pixel. Moreover, as we found, the surface leakage current strongly depends on the cathode bias, and the total leakage current turns out to be more than the sum of two separately measured bulk and surface components.

In this paper we report on these effects studied using our CZT pixel detectors coupled to custom VLSI chips. The ability of the readout system to simultaneously read signals from multiple pixel events was a key advantage in these studies.

## II. EXPERIMENTAL SETUP

The CZT detectors used in these studies were grown with the high-pressure Bridgman technique and patterned according to our specifications at eV-Products, Inc. Each detector had a pixel array with a grid of steering strips on the anode side (see Fig. 1), and a monolithic contact as the cathode. Geometrical parameters of the detectors and pixel patterns are listed in Table 1. Pt contacts were sputtered on the polished and chemically treated surface of CZT slabs and, as a final step, surface areas between contacts were passivated based on eV-Products, Inc. technology. To the best of our knowledge, the passivation was achieved by oxidation of the top layer of the CZT surface (we will discuss this in more detail later). In the course of the project we accumulated a great deal of data by characterizing detectors fabricated over a three-year period from different CZT ingots.

*I-V* characteristics of leakage currents were measured using a probe stage with a Keithley 237 SourceMeter and an EDC 521 voltage source. The detector was placed on a massive copper chuck attached to a cooling system, and the whole probe stage was enclosed inside a metal light-tight box. We took the majority of the data at room temperature; we also measured the temperature dependency of leakage currents at several specific biases by varying slowly the chuck temperature. Dry nitrogen gas was used to purge the box to avoid moisture condensation at low temperatures. The temperature during the measurements was monitored with a thermocouple (accuracy +/-0.1 C) attached to the chuck in close proximity to the detector.

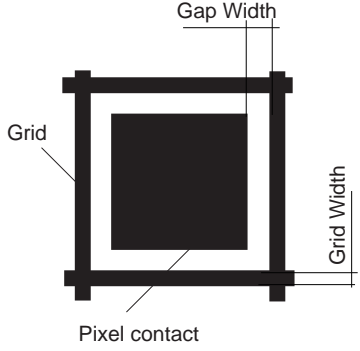

Fig. 1. A pixel contact enclosed inside a steering grid.

TABLE I

| Detectors | D1 | D2 | D3 |
|---|---|---|---|
| Thickness | 2 mm | 2 mm | 1.7 mm |
| Pixel pitch | 650x680 µm | 8x8 mm | 400x400 µm |
| Gap width | 100 µm | 200 µm | 50 µm |
| Grid width | 50 µm | 2 mm | 15 µm |

We took special precautions to ensure a steady state current condition during the measurements. Because of the presence of deep levels in the forbidden gap of CZT material, it can take several minutes or even hours to reach equilibrium between free and trapped carriers. To reduce the waiting interval between measurements, we varied the bias on the contacts in small steps. To be more specific, after setting a new set of contact biases, we paused for 5-10 minutes before taking 100 sequential measurements of the current, separated in time by 10-20-sec intervals. This sequence of data points allows us to verify that equilibrium has been actually achieved, and also increases the accuracy of the measurements. We found that illumination of the CZT slab with infrared (IR) light helps to reach equilibrium faster, especially in the case of surface current measurements. For this purpose (and some others which will be discussed later) we install a conventional IR light-emitting diode (with a spectral maximum at ~900 nm) in close proximity to the CZT sample.

We typically took the measurements from -100 V to +100 V, but in some cases we increased the maximum applied voltage up to 1 kV. We eliminate the leakage current flowing over the side surfaces of the detector by using a guard ring.

More details about the system and the measurement procedure can be found in Ref. [10].

## II. RESULTS OF THE LEAKAGE CURRENT MEASUREMENTS

We measured the bulk and surface leakage currents for individual pixels. In the case of the bulk current measurements, for each cathode bias the current flowing through the pixel was measured. The grid was kept under ground potential to ensure that the whole anode surface of the detector was at the same ground potential as the pixel contact (at a virtual ground); i.e. the electric field is uniform along the anode.

In the case of the surface current measurements, the grid voltage was varied in steps, while the cathode was kept at constant potential set with an additional power supply.

It should be mentioned, that although the detectors have different geometrical parameters, we found that results were very consistent after scaling the applied voltages to the same width of the gap (in the case of the surface current measurements) or to the same thickness of CZT (in the case of bulk current measurements).

### A. Bulk leakage current

Figs. 2 and 3 show two selected sets of *I-V* measurements whose shapes can be attributed to diode-like and ohmic-like behaviors; the majority of the measured *I-V* curves lie in-between those two. At first glance, these two curves look different; nevertheless, they can both be explained based on the same model which considers a CZT detector as a metal-semiconductor-metal (MSM) system with two back-to-back Schottky barriers (see e.g. Sze et al. [11] and Cisneros et al. [12]). In order to obtain good quantitative agreement with experimental data, two additional CZT features hove to be included in the model: series resistance of the bulk, and the effect of the interfacial layer between the metal contact and semiconductor. The latter effect can be described based on a so-called combined interfacial layer-thermionic-diffusion (ITD) model developed by Wu [13]. All the computational details can be found in Ref. [10]; here we present the final results of the analysis with a general explanation of the physical process involved.

The results of the fitting procedure are also shown in Fig. 2 and 3. At small applied biases the current follows Ohm's law simply because the resistance of the Schottky contact is much less than the series resistance of the undepleted bulk. The voltage drop across the Schottky barrier is small, and adjusts itself to accommodate the current limited by the bulk resistance. Above ~1 V the voltage drop across the barrier rises and current flowing over the barrier approaches the saturation limit; from this point on, bulk resistance has little effect on the current, and the width of the depletion layer is adjusted in accordance to the current. Above the *reach-through* voltage, $V_{RT}$, i.e. the voltage at which the detector becomes fully depleted, the dark current is controlled solely by the properties of the contact. The specific shape of the *I-V* curve below $V_{RT}$ can be calculated numerically if the parameters of the contact barrier are known. One can also solve an inverse problem–fit experimental data to estimate the unknown parameters used in the model. By doing so, we found, for example, that the height of the potential barrier of the Pt-CZT system lies between 0.77 and 0.79 eV for all tested detectors. Above $V_{RT}$ an analytical expression can be used to fit the measured curves (again, see Ref. [10] for details). The role of the interfacial layer, whose origin will be discussed later, is twofold. First, it affects the carrier concentration near the cathode (we assume n-type CZT) since the electrons entering the semiconductor from the metal should tunnel through the insulating layer. In ITD theory, this is described with the transmission coefficient $\theta$, which gives the fraction of electrons passing through the layer. Second, the voltage drop across the interfacial layer results in barrier height lowering which in turn leads to an exponential rise of the saturation current at high biases, above $V_{RT}$ (this is clearly seen on all $I$-$V$ curves above 200 V). Even though the barrier lowering effect can be neglected, the saturation current can still increase with the applied voltage depending on the conditions occurring near the cathode. If the electrons at the cathode are in equilibrium with the metal (competition between electrons crossing the interfacial layer into CZT and electrons being removed from the interface by the electric field toward the anode), then the free carrier concentration at the cathode remains independent of the applied voltage up to the very high biases. In this case the saturation current is simply proportional to the electron concentration and electric field-strength at the cathode; in the Schottky-depletion layer approximation the field is proportional to the square root of the applied voltage below $V_{RT}$ and directly proportional to the voltage above $V_{RT}$. This is called the diffusion-limited approximation of the Schottky barrier. The other limiting case is when the electrons entering the CZT are immediately removed from the contact area. In this case the current will be determined by the flux of electrons thermionically emitted over the potential barrier from the metal into the CZT. This current, called thermionic emission-limited current, depends only on the height of the barrier, and is independent of the applied voltage. In reality, we observed intermediate cases where the ITD theory will be a more proper approach. The effects described above are clearly seen on the measured $I$-$V$ curves. As an example, the $I$-$V$ curve in Fig. 2 exhibits a very strong effect of the interfacial layer. Since the transmission coefficient is much less than 1, the saturation current is close to the thermionic-limited case. The $I$-$V$ curve starts bending around 1 V, and the density of the dark current is two orders of magnitude less than the thermionic emission-limited current that would be expected based on the potential barrier height of ~0.78 eV. However, above 100 V it starts rising exponentially because of the potential barrier lowering effect. If the latter effect could be neglected we would expect to see I-V characteristic following the curve "$T$" in Fig. 2 (the thermionic emission-limited current). Moreover, if we could also neglect electron reflection from the interfacial layer, i.e. set $\theta=1$, we would expect the current to follow the curve "$D$" (diffusion-limited current). As shown in Fig. 2, the measured current is closer to the thermionic-limited case. The curve evaluated based on the ITD theory is also plotted along with estimates for the model parameters. This curve provides an excellent fit to the experimental data. The fact that the measured current is close to the thermionic-limited case (if one neglects the barrier lowering effect) is directly related to the poor electron transmission across the interfacial layer, i.e. $\theta<<1$.

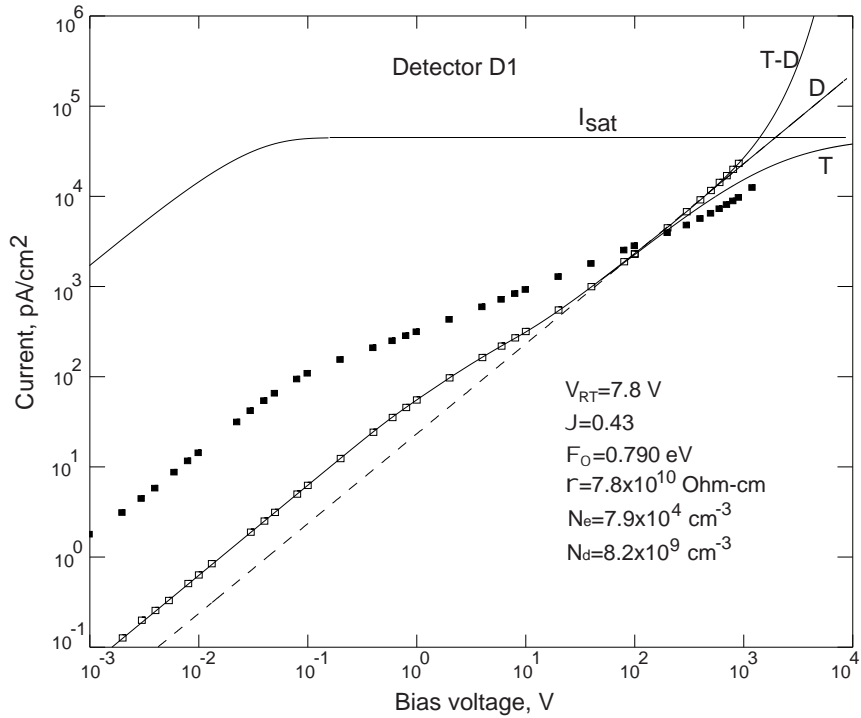

Fig. 2. *I-V* characteristics of the bulk current measured with (solid squares) and without (open squares) IR- light illumination. "T-D" is the result of fitting the dark current using the ITD theory of the Schottky-barrier contact. Curves "D" and "T" are calculated in the diffusion and thermionic emission limits, respectively, and neglecting the potential barrier lowering effect (all the parameters are taken from the previous fit). $I_{sat}$ is the *I-V* characteristic of the ideal diode. The dashed line is the extrapolation of the calculated diffusion limited current to zero bias. The listed fitting parameters are: temperature, $T$; reach-through voltage, $V_{RT}$; transmission coefficient, $\theta$; potential barrier height, $\Phi_b$; bulk resistivity, $\rho$; free carrier concentration, $N_e$; and density of the ionized charges (space-charge density), $N_d$.

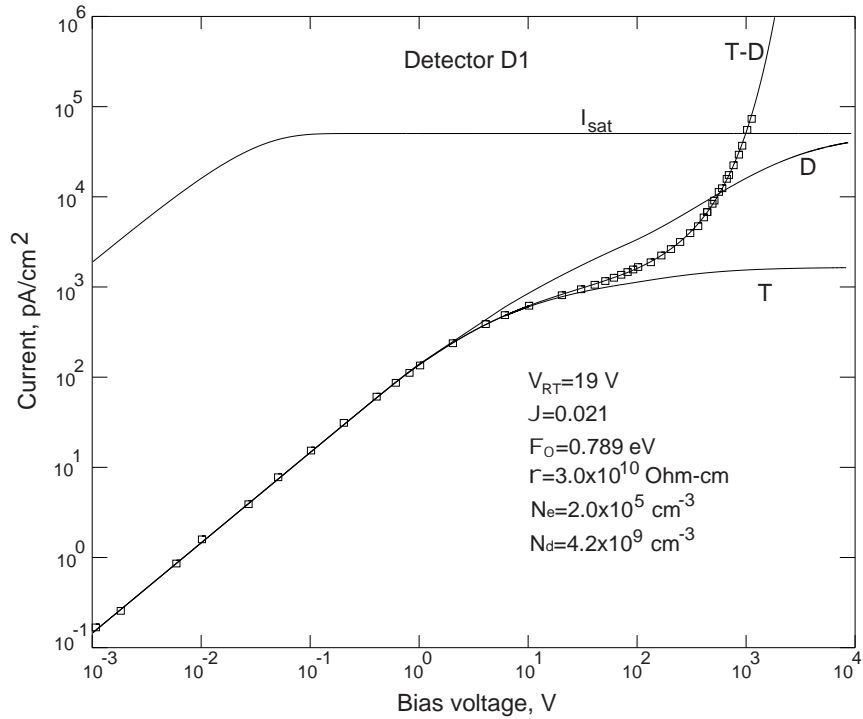

Fig. 3. Same as Fig. 2 but measured for a different CZT sample (no IR light illumination was used).

The curves in Fig. 3 represent another limiting case in which the effect of the interfacial layer is small. The fitting procedure gives similar values for the model parameters as in the previous case, except for the transmission coefficient $\theta$, which is now close to 100%. As a result, the measured current follows the diffusion-limited curve "$D$" up to very high biases, and the exponential rise due to barrier lowering is less pronounced. As mentioned previously, the diffusion-limited current increases linearly with voltage above $V_{RT}$. This behavior resembles Ohm's law, but with a much smaller *effective* resistivity which is inversely proportional to the free electron concentration at the cathode. In the diffusion approximation (see for instance Ref. [12]), the electron concentration at the cathode is less than in the bulk by a factor $exp(-V_{bi}/kT)$, where $V_{bi}$ is the built-in voltage, and $kT$ is the thermal energy at temperature $T$. Our fitting algorithm gives 0.03-0.05 eV for $V_{bi}$, then the reduction factor is ~0.1. In the bias range from 10 to 200 V, it is very easy to misinterpret the $I$-$V$ curve as following Ohm's law. As a result, an overestimate of the bulk resistivity will be obtained (the dotted line in Fig. 3 is the extrapolation of the diffusion-limited current to zero bias).

Based on the above discussion, we can conclude that the diode-like behavior of the Pt-CZT system is masked by other effects. It manifests itself in the fact that the actual measured leakage current is significantly smaller than it would be, based on Ohm's law, and only a vigorous analysis as described in Ref. [10] can reveal the parameters of the Schottky barrier. This is especially true for the curve in Fig. 3 which, in fact, represents more commonly observed $I$-$V$ characteristics. Here, slight changes of the slope can be barely seen. However, several simple experiments can uncover the diode-like behavior of the Pt–CZT interface.

One of them is to use IR light to illuminate the detector during the measurements; this effectively reduces the bulk resistivity of the material. The $I$-$V$ curve shown on the top of Fig. 3 was measured under this condition for the same detector. The sharp bending around 0.1 V, which is an indication of the diode-like behavior, is clearly seen (for comparison the ideal $I$-$V$ curve of a diode is also plotted).

Another experiment is to measure the current flowing between the contacts having different areas. When the grid was left floating we observed a strong asymmetry in the currents measured for two different polarities of the cathode bias. At a negative bias we measured a much higher current because the electrons were injected from the whole cathode (whose area is much larger than that of a pixel contact) and collected by a single pixel. When the bias was reversed, the electrons were injected from the small area pixel contact–as a result we measured a smaller current. This experiment proves that the CZT detectors used in these measurements were n-type.

Based on the above discussion we can conclude that in most cases, the bulk leakage current is controlled by the height of the Schottky barrier at the contacts (0.77-0.79 eV). In the first approximation, the shape of the $I$-$V$ curve resembles Ohm's law but with much smaller *effective* resistivity than in the bulk. However, if the interfacial oxide layer exists at the contacts, it changes the shape of the *I-V* curve toward the thermionic emission case. This can be used to characterize the quality of the contacts on CZT.

### B. Surface leakage current

Although the importance of surface effects on the performance of CZT detectors has been commonly accepted, the information about the properties of the CZT surface, and the origin of the high surface conductivity, in particular, is very limited.

The surface leakage current is apparently due to the presence of a low-resistivity surface layer which has different, in comparison with the bulk material, chemical and band structure. As is well know in semiconductor science, non-stoichiometric material is typically formed on the surface of semiconductors after the dicing and polishing steps. In many cases this non-stoichimetric layer can be removed by chemical etching; however, in the case of CZT, it appears that bromine etching leaves a Te-enriched layer on the CZT surface [14-17]. There are no direct measurements of the electronic properties of the Te layer on the CZT surface, but some interesting results were obtained for CdTe material. Montgomery [18] demonstrated that a thick layer of amorphous tellurium could be chemically produced on the surface of CdTe, and that this layer indeed had a very low resistivity because of the narrow band gap of Te, ~0.33 eV.

Our results, which we present below, support this general idea [14-17] that the Te layer is formed during the fabrication process, which later becomes partially or completely oxidized after the passivation step, or as a result of natural oxidation in air. The later process starts as soon as the surface of CZT is exposed to air or to the oxygen outgasing in the sputtering chamber. If for any reason the surface of CZT was not protected from oxygen prior to making the contacts, the insulating layer, whose effects on the bulk leakage current were discussed above, will appear between the contact and CZT surface.

Despite the surface oxidation, there always remains a stoichiometrically imperfect interfacial layer between the native oxide (Te oxide) and the bulk material. We believe that the properties of this layer, which can be considered as an example of the so-called insulator-semiconductor interface (see for instance Ref. [19]) determine the surface conductance of CZT detectors. The main feature of the interfacial layer is a very high concentration of states (associated with different kinds of defects, impurities, Te clusters, etc) located in the forbidden gap of CZT, which can form a band structure. If the band gap in the interfacial layer is small, the charges from the semiconductor can easily occupy conduction or valence bands, and make possible a direct-current conduction.

Fig. 4 shows two *I-V* curves measured for the detector D2 at two temperatures: 26 C and –10 C. As seen in both cases, the current changes first as a linear function of the voltage, applied across the gap between the grid and the contact, up to a certain transition voltage, above which it starts rising rapidly as a power law function, $I \sim V^\alpha$, with $\alpha > 2$. More accurate measurements reveal, however, that the dependence is linear only within close proximity of zero bias. At higher biases, but below the transition voltage, it rises just a little slower than a linear function. This fact, together with the asymmetrical *I-V* curves observed for the pixels with small contacts and wide gaps, i.e. when the perimeter of the grid is much larger than the perimeter of the contact, suggests that the conducting layer on the CZT surface acts as the MSM system described above. The transition voltage (or voltage of crossover from Ohm's law to power law) is related to a breakdown condition at the contact, but the current does not rise sharply because the space charge accumulated in the gap limits current injection from the contact. Moreover, the space-charge-current regime predicts a power law dependence with $\alpha > 2$ for semiconductors with an exponential distribution of traps in the forbidden gap (see e.g. Lampert and Mark [20]).

The existence of the space-charge-limited regime is an indication that surface conductance takes place inside the thin interface layer, rather than inside the accumulation or inversion layers that generally can exist in the bulk near the surface. The different electronic properties of the bulk and surface layer can be illustrated by the temperature dependence of the surface and bulk currents. Fig. 5 shows the relative changes in the bulk and surface currents measured for the detector D2. The curve (1) represents the surface current measured at 1 V bias on the grid and zero bias on the cathode, whereas the curve (2) represents the bulk current measured at 1 V on the cathode and zero bias on the grid. (The elected voltages correspond to the linear regions of the bulk and surface *I-V* characteristics.)

We found, that the temperature dependence of the bulk leakage current, measured at –1 V, can be accurately fitted with a single activation energy function with $E_a$=0.84 eV. It is tempting to attribute this energy to a deep donor level at 0.743 eV above the valence band which is, according to Ref. [21], responsible for compensation in CZT crystals grown by the high-pressure Bridgman method. In the case of the surface leakage current, a similar fit with the single activation energy function gave $E_a$=0.65 eV.

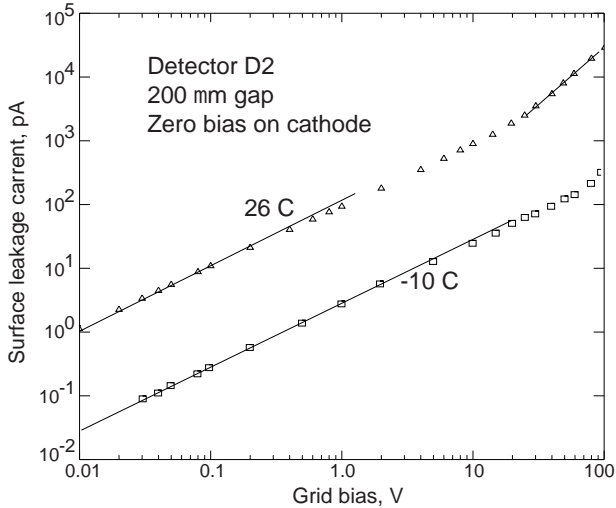

Fig. 4. Surface leakage currents measured for the same CZT detector at 26 C (triangles) and at –10 C (squares). Negative biases were applied on the grid; the cathode was under ground potential.

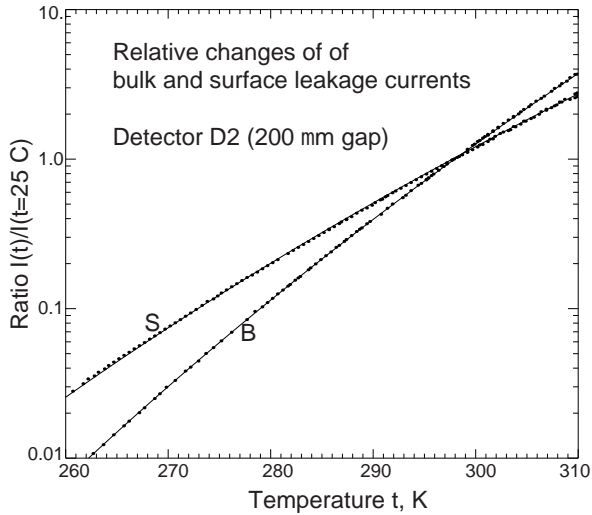

Fig. 5. Relative changes of the bulk (B) and surface (S) currents with temperature. The bulk current was measured with –1 V on the cathode and zero bias at the grid. The surface current was measured with –1 V on the grid and zero bias at the cathode.

We estimated the surface resistivity by fitting the measured $I$-$V$ curves in the low-voltage region. Normally, we accepted from the vendor only high surface resistivity detectors, with resistivity >500 GOhm/square. Several treatments can increase the surface resistivity even more. All these treatments can be explained by the surface oxidation, which reduces the number of the interface states and consequently the free carrier concentration inside the interfacial layer. Those are: 1) natural surface oxidation over a long period of time when detectors are kept in air; 2) temperature stimulated oxidation (baking the detectors in air); and 3) electric field stimulated oxidation (keeping the detectors biased). Usually the surface resistivity increases 2-5 times, depending on its initial value, and stabilizes around 2000-3000 GOhm/square. The highest surface resistivity we measured was ~4500 GOhm/square at room temperature.

The shape of the $I$-$V$ characteristic of the surface current helps to characterize the quality of the surface passivation; however, they cannot be used to estimate the total pixel leakage current because the cathode bias significantly affects the surface conductance.

### C. The field-effect

The field-effect is used in many semiconductor devices, e.g. a field-effect transistor, where the width of the space-charge layer near the surface is controlled by the external electric field created with the metal gate contact separated from the semiconductor by a thin insulating layer. We attempted to observe a similar effect in CZT detectors by placing a thin one-sided aluminized mylar foil on top of the contact pattern of the detector. We did not observe any changes in the surface leakage current (even when the applied voltage exceeded 100 V). This implies that the surface conductance of CZT is not due to accumulation or inversion layers.

It is possible that there could be a different type of field-effect in the CZT detectors as described below. Assuming that surface leakage current is flowing inside the insulator-semiconductor interface, one can expect that its conductance can be controlled by changing the carrier concentration inside the interface (i.e. by changing the occupancy of the conduction or valence bands). To check this hypothesis, we tried to observe the changes in surface leakage current caused by applying voltage on the cathode. Fig. 6 shows the $I$-$V$ characteristics of the surface current measured at different biases on the cathode. Before plotting the $I$-$V$ curves, we subtracted the bulk current components by fitting the $I$-$V$ curves in the low bias region. As is seen in Fig. 6, the surface current changed significantly when voltage was applied on the cathode. This means that the total pixel leakage current does not equal the sum of the bulk and surface currents if they were measured separately. We tested a large number of detectors and found that when a negative voltage was applied on the cathode, the actual total pixel leakage current was always significantly larger than that calculated by adding the bulk and surface components. In contrast, when the cathode was positively biased the actual measured current was smaller. This is a disappointing result, since the negative bias on the cathode (the one employed for normal operation of a pixel detector) increases the surface leakage current. To illustrate this, Fig. 7 shows the bias dependence of the surface conductance, evaluated from plots similar to those shown in Fig. 6. We systematically tested all our detectors (~20) and found one unique detector for which the polarity of the field-effect was "reversed" (i.e. the surface leakage current was significantly smaller when negative bias was applied on the cathode). We are currently investigating this effect, which is of great importance for multi-electrode CZT detectors, in cooperation with eV-Products, Inc.

Comparing the plots in Figs. 4 and 6, one can easily see the similarities in the way the surface $I$-$V$ curves evolve when the temperature decreases, and when a positive voltage is applied on the cathode. In both cases the ohmic region extends toward the higher grid biases, while the non-ohmic part of the curve rises with a steeper slope. Similar changes were observed after baking the samples or keeping the samples in air for a long time (i.e. after further oxidation of the CZT surface). All these effects can be explained as a result of the reduction of free carriers in the insulator-semiconductor interface. The electrons excited in the conduction bands of the interface states remain in equilibrium with the carriers in the bulk; as a result, the occupancy of the conduction bands of the interface varies with the carrier concentration in the bulk near the surface. If the latter is reduced (because of the cathode voltage or low temperature), the former should also decrease as shown in Figs. 4 and 6.

## III. EFFECTS OF SURFACE AND BULK LEAKAGE CURRENTS

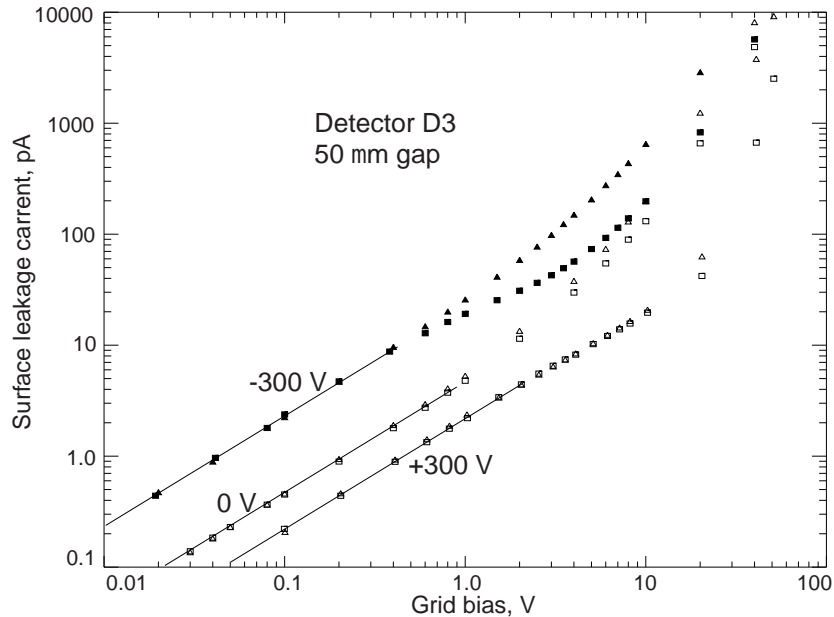

Fig. 6. *I-V* characteristics of the surface current measured at three different biases on the cathode at room temperature. The triangles and squares represent current measured at negative and positive biases on the grid respectively. The solid lines are the fitting results at low biases with a linear function. The bulk current components are subtracted to adjust the curves to zero current at zero bias.

The performance of the CZT pixel detector is directly related to the dark current and surface conductivity of the detector. The total leakage current determines the electronic noise. Moreover, since the signals from the pixels are read out with a VLSI chip bonded directly to the detector, it is desirable, from a designer's point of view, that the leakage current remains low. This means that the pixel area should be small or the operating temperature low.

Because of the surface conductance, the field lines in the detector can intersect the surface between the contacts where the charge gets trapped. As has been already discussed, to avoid this problem we have the steering grid negatively biased with respect to the pixel contacts. This adds the surface component to the total leakage current which, however, can be reduced by proper treatment of the surface.

### A. Total leakage current per pixel

Because of the field-effect the total leakage current per pixel should be measured by applying bias simultaneously on the cathode and the grid. The results are given here for the D3 detector, whose geometrical parameters: a 400x400 μm pitch size, a 50 μm gap, and a 15 μm grid are very close to the optimal. For a −5 V bias on the grid and −300 V on the cathode (the typical operating biases), the total leakage current was measured to be ~150 pA per pixel at 26 C; it drops below 20 pA at 0 C and below 3 pA at −10 C. These are typical currents measured for all but one detector, which has a different polarity of the field-effect (see a previous section). Fig. 8 shows how the total current measured for this detector at

−6 V on the grid changes with the cathode bias. At 26 C and −300 V on the cathode, it drops below 7 pA, a value limited by the bulk leakage current only. This unique detector is an example of a truly room temperature detector which, as we hope, eV-Products, Inc. will be able to reproduce in the future.

As was explained earlier, at high grid biases the surface leakage current can reach the space-charge-controlled regime and become very large. However, statistical fluctuation of the space-charge-limited current (as is well known in semiconductor device physics) remains small. This may explain why the electronic noise of the coplanar CZT detectors was found to be significantly less than the noise estimates based on the nominal value of the leakage current [22].

### B. Charge collection efficiency

In the absence of surface conductivity all the electric field lines are expected to originate and terminate on the metal contacts only. This is because the CZT is slightly conducting, and charges built up on the surface in the areas between the contacts repel field lines toward the contacts. In this ideal case, field lines should never intersect non-metalazed surfaces. However, if the surface is conductive the built-up charge leaks away and field lines can intersect the surface areas between the contacts. This effect exists in any multi-electrode CZT detector and should be taken into account when designing a contact pattern.

Another important feature related to charge collection efficiency is the non-uniformity of the electric field over the detector thickness, which comes about because of the

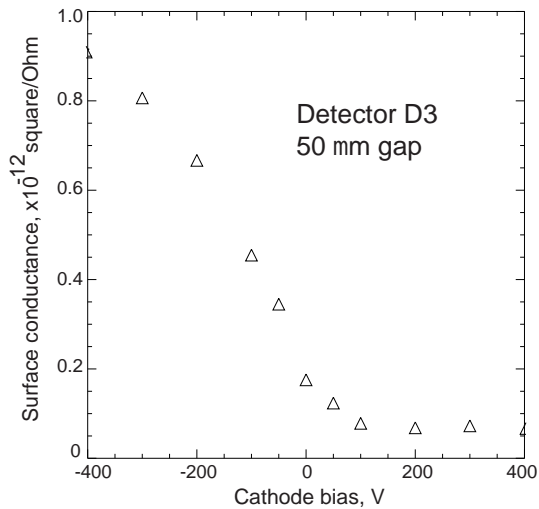

Fig. 7. Surface conductance versus cathode bias.

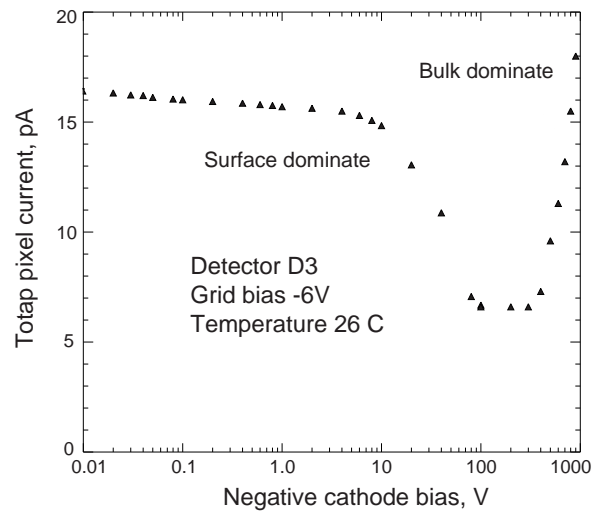

Fig. 8. Total leakage current per pixel measured for the "unique" detector (see text) at −6 V on the grid and negative biases on the cathode. The surface leakage current measured at −200 V on he cathode is three times less than that at zero bias.

Schottky-barrier contacts. In the oversimplified case of uniform distribution of space charge, the electric field strength decreases linearly toward the anode (for a n-type material). In reality, because of the deep levels, the field seems to decrease more rapidly, especially at low temperatures.

## IV. CONCLUSION

These studies demonstrate the importance of the dark current and surface conductance for CZT pixel detector. Both bulk and surface leakage currents contribute to the total leakage current per pixel, which determines the electronic noise, and limits energy resolution of the detector. They also control the field distribution inside the detector and, consequently, charge collection efficiency.

We found that the bulk leakage current is controlled by the height of the Schottky barrier, which is about 0.77-0.79 eV. The shape of the *I-V* curve resembles an ohmic-like dependence but with an order of magnitude less effective resistivity than in the bulk. The interfacial oxide layer, which is inevitably formed between the contact and CZT surface, affects the shape of the *I-V* characteristic, and this can be used to characterize the quality of the contacts.

To avoid the charge loss at the surface between the pixel contacts, we employ a very thin grid between the contacts to steer electrons landing on the surface toward the contacts. It is very important that the width of the grid and the gap between the grid and the contact be small (we suggest a 15 um grid and a 50 um gap). In this case the required grid bias will be below the region where the surface space-charge-controlled current regime takes place. Because of the field-effect the total leakage current per pixel becomes a complex function of the applied biases. The actual current can be significantly higher or significantly lower than the sum of the bulk and surface currents measured separately, depending on surface properties. Needless to say, this effect (similar to the field-effect in semiconductor devices) should also play an important role in coplanar grid and other multi-electrode CZT detectors.

## V. ACKNOWLEDGEMENT

This work was supported by NASA under grant No. NAG5-5289. The authors wish to thank K. Parnham and C. Szeles from eV-Products, Inc. for fruitful discussions.